\documentclass[prl,twocolumn,amsfonts,noshowpacs,nofootinbib,amsmath,amssymb]{revtex4}

\usepackage[dvips]{graphicx}
\usepackage{amsmath,amssymb}
\def\mr{\mathrm}
\def\mb{\boldsymbol}
\def\mc{\mathcal}

\newcommand{\maru}[1]{\left( #1 \right)}

\newcommand{\bea}{\begin{eqnarray}}  \newcommand{\eea}{\end{eqnarray}}
\newcommand{\beq}{\begin{equation}}  \newcommand{\eeq}{\end{equation}}
  
\newcommand{\lmk}{\left(}  \newcommand{\rmk}{\right)}
\newcommand{\lkk}{\left[}  \newcommand{\rkk}{\right]}

 \newcommand{\calP}{{\cal P}}
\newcommand{\msolar}{M_\odot}
\newcommand{\Psik}{\Psi_{\mb{k}}}
\begin{document}

%\title{Gravitational-Wave Constraints on Abundance of Primordial Black Holes}
\title{Gravitational wave background
 as a probe of the primordial black hole
abundance}
\author{Ryo Saito$^{1,2}$} \author{Jun'ichi Yokoyama$^{2,3}$}
\affiliation{$^1$Department of Physics, Graduate School of Science,  
The University of Tokyo, Tokyo 113-0033, Japan\\
$^2$Research Center for the Early Universe (RESCEU),
Graduate School of Science, The University of Tokyo,Tokyo 113-0033, Japan\\
$^3$Institute for the Physics and Mathematics of the Universe, University of Tokyo, Chiba 277-8568, Japan
}

\begin{abstract}
Formation of significant number of primordial black holes (PBHs)
is realized if and only if primordial density fluctuations have a
large amplitude, which means that tensor perturbations generated 
from these scalar perturbations as a second order effect are also
 large and comparable to the observational data.
We show that pulsar timing observation could find/rule out 
PBHs with $\sim 10^{2}\msolar$ which are considered as
a candidate of intermediate-mass black holes and that PBHs with
mass range $10^{20-26}$ g, which serves as a candidate of dark matter, 
may be probed by future space-based laser interferometers and atomic interferometers. 
\end{abstract}

\maketitle

%%%%%%%%%%%%%%%%%%%%%%%%%%%%%%%%%
%Introduction
%%%%%%%%%%%%%%%%%%%%%%%%%%%%%%%%%
%\section{Introduction}\label{s:intro}

Primordial black holes (PBHs) are produced when  density fluctuations 
with a large amplitude enters the horizon in the 
radiation dominated stage of the early universe 
with their typical mass given by the horizon mass 
at that epoch \cite{PBH,PBHreview}.
%	\begin{equation}\label{eq:mass}
%		M_\mr{PBH} = \frac{4\pi}{3}\rho(H^{-1})^3 = \frac{1}{2GH},
%	\end{equation}
PBHs with their mass smaller than $10^{15}$g would have been evaporated
away by now due to Hawking radiation \cite{Hawking:1974rv}. The abundance of 
these light holes has been constrained by big-bang nucleosynthesis (BBN) 
\cite{Kohri:1999ex}
and gamma-ray background \cite{gammaray} etc.

 Heavier PBHs,
on the other hand, can play some astrophysical
roles today.  For example, they may serve as an origin of the
intermediate-mass black holes (IMBHs), which are considered to be the 
observed ultra-luminous X-ray sources, if their mass and abundance
lie in the range
 $M_\mr{PBH} \sim 10^2 M_{\odot}-10^4 M_{\odot}$
and $\Omega_\mr{PBH}h^2 \sim 10^{-5} - 10^{-2}$, respectively
\cite{Kawaguchi:2007fz}. PBHs with mass 
$M_\mr{PBH} \sim 10^{20}\mr{g}-10^{26}\mr{g}~
(10^{-13}M_{\odot}-10^{-7}M_{\odot})$ \cite{PBHreview,
Abramowicz:2008df}
 and the abundance
$\Omega_\mr{PBH}h^2 = 0.1$ \cite{Komatsu:2008hk}
can provide an astrophysical 
origin of dark matter (DM) which satisfies the constraint
imposed by gravitational lensing experiments \cite{Marani:1998sh,DM}.
%\footnote{In \cite{Seto:2007kj}, it is 
%discussed that the pulsar timing observations 
%can constrain the abundance of the PBHs with 
%mass $M_\mr{PBH} \sim 10^{25}\mr{g}$.}.

Formation of the relevant number of PBHs on a specific mass scale
is realized if the power spectrum of primordial density fluctuations
has a peak with amplitude  $10^{-2}-10^{-1}$ 
on the corresponding scales (See \cite{models,jy} for inflation 
models to realize such spectra).
In such a situation the second-order effects are expected to play
an important role.  For example,
they generate non-Gaussianity in
 the statistical distribution of density fluctuation, and the amount of PBH production
could be modified \cite{NG0}. Such an effect was recently investigated in single-field inflation models,
but it turned out that the non-Gaussian effect is negligibly 
small \cite{NG}, 
justifying previous analysis assuming Gaussianity \cite{Carr:1975qj}.

Second-order effects also generate
tensor fluctuations to produce stochastic 
background of gravitational waves (GWs) 
from scalar-tensor mode coupling 
 \cite{Ananda:2006af,Baumann:2007zm}. Their amplitude may well 
exceed the first-order tensor perturbation generated by quantum
effect during inflation \cite{starogw} in the current set up 
since the amplitude of density 
fluctuations required for a substantial density of PBHs is so 
large. Furthermore, the amplitude is expected to exceed that of GWs 
generated 
during the PBH collapses since smaller amplitude of density fluctuations 
suffices to produce the second-order GWs with a relevant amplitude, 
compared to those 
necessary for the formation of PBHs.

 In this Letter, we show the GWs induced by scalar fluctuations
as a second-order effect \cite{Ananda:2006af,Baumann:2007zm} 
is a useful probe 
to investigate the abundance of the PBHs.  We calculate the spectrum
of these second-order GWs in the case that scalar fluctuations have
a sufficiently large peak to realize the formation of 
appreciable numbers of PBHs.  As a natural consequence, we 
find that the spectrum of GWs has a peak on a scale approximately
equal to the scale of the peak of the scalar fluctuations.
We can therefore obtain information on the abundance of PBHs
with the horizon mass when the scale of the peak entered the
Hubble radius by observing GWs with the frequency corresponding
to the same comoving scale, namely, 
$10^{-10}\mr{Hz}-10^{-9}\mr{Hz}$ for the IMBHs produced primordially and 
$10^{-5}\mr{Hz}-10^{-2}\mr{Hz}$ for the dark-matter PBHs. Fortunately, 
the former band can be probed by 
 the pulsar timing observations 
\cite{Thorsett:1996dr,Jenet:2006sv} while the 
latter band can be observed in the future by 
space-based laser interferometers \cite{LISA, DECIGO, BBO} 
as well as atomic gravitational wave interferometric sensors (AGISs)
 \cite{Dimopoulos:2008sv} for the dark-matter PBHs.
  
We write the perturbed metric as
\begin{equation}\label{eq:metric}
\mathrm{d}s^2 = a(\eta)^2 \left[ -e^{2\Phi} \mathrm{d}\eta^2 
+ e^{-2\Psi}(\delta_{ij}+h_{ij})\mathrm{d}x^i \mathrm{d}x^j \right], 
\end{equation}
including both scalar perturbations, $\Phi$ and $\Psi$, and
tensor perturbation, $h_{ij}$, which satisfies 
$\partial_{i}h^{i}_{j}=h^i_i=0$ 
with $h^{i}_{j} \equiv \delta^{ik}h_{kj}$. 
We assume the lowest-order tensor perturbations are negligible and incorporate only those 
generated by the scalar mode as a second-order effect.
The relevant part of the second-order Einstein equation therefore reads 
\begin{equation}\label{eq:tee}
{h^{i}_{j}}''+2\mathcal{H}{h^{i}_{j}}'-\partial^2h^{i}_{j }
=2\mathcal{P}^{is}_{rj}S^{r}_{s},
\end{equation}
where a prime denotes differentiation with respect to the conformal
time, $\eta$, $\mathcal{P}^{is}_{rj}$ represents the
 projection operator to the transverse, traceless part, 
and $\mathcal{H}\equiv a'/a$ 
\cite{Ananda:2006af,Baumann:2007zm}. Here, the source term reads
\begin{equation}\label{eq:source}
S^{r}_{s} = -2\Psi\partial_{r}\partial_{s}\Psi 
+\frac{4}{3(1+w)}\partial^{r}(\Psi+\mathcal{H}^{-1}\Psi')
\partial_{s}(\Psi+\mathcal{H}^{-1}\Psi'),
\end{equation}
with $w \equiv \rho/p$ being the equation-of-state parameter of the
background fluid. In practice, only the radiation dominated era is
relevant, so we take $w=1/3$ hereafter. 
We also neglect anisotropic stress, which is
expected to give only a small correction \cite{Baumann:2007zm},
and set $\Phi=\Psi$ at linear order. Note the source term is second-order with 
respect to the scalar perturbations and absent at linear order.
In order to calculate the induced GWs up to second order, therefore, 
it is sufficient to use the linear scalar modes. 
Hence, we only need to solve the linear evolution 
equation \cite{Mukhanov},
\begin{equation}\label{eq:see}
\Psik''(\eta)+\frac{4}{\eta}\Psik'(\eta)+ \frac{k^2}{3}\Psik(\eta) =0,
\end{equation}
for the scalar modes, where $\Psik$ represents a Fourier mode of $\Psi$. 
Its non-decaying solution is given by
$ \Psik(\eta)=D_k(\eta)\Psik(0)$ with the transfer function
\beq
D_k(\eta)=\frac{3}{(k\eta)^2}\lkk\frac{\sqrt{3}}{k\eta}
\sin\lmk\frac{k\eta}{\sqrt{3}}\rmk - \cos\lmk\frac{k\eta}{\sqrt{3}}\rmk
\rkk. \label{transfer}
\eeq

For our purpose we assume the form of the power spectrum
of the initial fluctuations to be approximated by the Dirac delta function 
with respect to $\ln(k)$,
\beq
  \calP_\Psi(k)\equiv \frac{k^3}{2\pi^2}\langle |\Psik(0)|^2\rangle 
= \mc{A}^2 \delta_D(\ln(k/k_p)), \label{spectrum}
\eeq
where $k_p$ and $\mc{A}^2$ represent  the wavenumber of the peak
and $(\text{amplitude})^2 \times \ln(\text{peak width})$ 
of the original spectrum, respectively.
With this power spectrum the fractional energy density of the
region collapsing into PBHs at their formation time is estimated as
\begin{equation}\label{eq:beta}
\beta(M_{\mr{PBH}}) \sim 0.1 \exp\maru{-\frac{\Psi_c^2}{2\mc{A}^2}},
\end{equation}
where $M_{\mr{PBH}}$ is of the order of the horizon mass when the
comoving scale $k_p^{-1}$ enters the Hubble radius and $\Psi_c$ is the
threshold value of PBH formation.  Carr \cite{Carr:1975qj} takes
the threshold value of the density contrast to be
$\delta_c=1/3$ corresponding to $\Psi_c=1/2$.  Analysis based on 
numerical calculation  \cite{shibatasasaki} gives a similar but 
slightly different value \cite{jy}.
One can express
the current value of the density parameter of PBHs in terms of
 $\beta(M_{\mr{PBH}})$ as
\begin{equation}
\Omega_{\mr{PBH},0}h^2=1\times 10^{14}\beta(M_{\mr{PBH}})
 \maru{\frac{M_{\mr{PBH}}}{10^{20}\, \mr{g}}}^{-1/2}
	\maru{ \frac{g_{\ast p}}{106.75} }^{-1/3},%\label{omegapbh}
\end{equation}
where $g_{\ast p}$ is the effective number of the relativistic 
degrees of freedom when the peak scale $k_p^{-1}$ entered the Hubble radius.

We define the Fourier modes $h_{\mb{k}}$ by
\begin{equation}%\label{eq:ftt}
h_{ij}(\mb{x},\eta) = \int\!\frac{\mr{d}^3k}{(2\pi)^{3/2}}e
^{i\mb{k}\cdot\mb{x}}\left[ h_{\mb{k}}^{+}(\eta)\mr{e}_{ij}^{+}(\mb{k})
 + h_{\mb{k}}^{\times}(\eta)\mr{e}_{ij}^{\times}(\mb{k}) \right],
\end{equation}
where $\mr{e}_{ij}^{+}(\mb{k}),\mr{e}_{ij}^{\times}(\mb{k})$ 
are polarization tensors which are normalized as 
$\sum_{i,j} \mr{e}_{ij}^{\alpha}(\mb{k})\mr{e}_{ij}^{\beta}(-\mb{k})
=2\delta^{\alpha\beta}$. 
The Fourier transform of the source term (\ref{eq:source}) is 
also defined similarly.  We find the source term is constant
when $k_p\eta/\sqrt{3} \ll 1$, while it decreases in proportion to
$\eta^{-2}$  for $k_p\eta/\sqrt{3} \gg 1$.  As a result
the production of scalar-induced GWs mostly occurs around the time
when the peak scale $k_p^{-1}$ crosses the sound horizon.  Using the Green function method 
one can easily find a formal solution to (\ref{eq:tee}), from which
we can evaluate the density parameter of GWs contributed by a 
logarithmic interval of the wavenumber around $k$. 
It is formally expressed as
\beq
\Omega_{\mr{GW}}(k,\eta) = \frac{k^3}{12\pi^2 \mc{H}^2}
\maru{ {{|h_{\mb{k}}^{+}}'|}^2 + |{{h_{\mb{k}}^{\times}}'}|^2 }.
\eeq
This is valid for modes well inside the horizon \cite{Maggiore:1999vm}.
In the radiation dominated regime it is explicitly given by
%%\begin{widetext}
\begin{align}
\Omega_{\mr{GW}}(k,\eta) &= \frac{2}{3}
\int^{\eta}\!\!\!\!\mr{d}\eta_1\int^{\eta}\!\!\!\!\mr{d}\eta_2~
\eta_1\eta_2 \label{eq:endbysp} \\
&\times\sin\lkk k(\eta-\eta_1) \rkk \sin\lkk k(\eta-\eta_2) \rkk
{\mc{S}_{\mb{k}}}(\eta_1,\eta_2), \nonumber
\end{align}
where we have defined
\begin{align}
\mc{S}_{\mb{k}}(\eta_1,\eta_2) &\equiv \int_{0}^{\infty}\!\!\!\!
\mathrm{d}\tilde{k}\int_{-1}^{1}\!\!\!\!\mathrm{d}\mu~
\frac{k^3\tilde{k}^3}{|\mb{k-\tilde{k}}|^3}(1-\mu^2)^2 \\ 
&\times f(\tilde{k},|\mb{k-\tilde{k}}|,\eta_1)
f(\tilde{k},|\mb{k-\tilde{k}}|,\eta_2) \nonumber\\
&\times
\mathcal{P}_{\Psi}(\tilde{k})\mathcal{P}_{\Psi}(|\mb{k-\tilde{k}}|).
\nonumber
\end{align}
%%\end{widetext}
Here $f(k_1,k_2,\eta)$ is a function written in terms of the transfer 
function for the scalar modes as follows:
\begin{align}
&f(k_1,k_2,\eta) \equiv 2 D_{k_1}(\eta)D_{k_2}(\eta) \\
&~~~+[D_{k_1}(\eta)
+\mathcal{H}^{-1}D_{k_1}'(\eta)]
[D_{k_2}(\eta)+\mathcal{H}^{-1}D_{k_2}'(\eta)]. \nonumber
\end{align}
%\end{widetext}
In the mass range of the PBHs of our interest, creation of
scalar-induced GWs is terminated well before matter-radiation equality time.
After that the energy density of GWs decreases in proportion to
$a^{-4}$.  As a result the amplitude of the spectrum $\Omega_{\mr{GW}}(f,\eta_0)$ 
at the peak frequency $f_{\mr{GW}} \equiv k_p/(\pi\sqrt{3}a_0)$ today
is given by 
%\begin{widetext}
\begin{align}
A_{\rm GW} &\equiv 6 \times 10^{-8}
\maru{\frac{g_{\ast p}}{106.75}}^{-1/3}
\!\!\!\maru{\frac{\mc{A}^2}{10^{-2}}}^2.
\label{eq:agw}
\end{align} 
%\end{widetext}
As expected, the amplitude of the induced GWs exceed
its first-order counterpart, $\Omega_{\mr{GW}}h^2 \sim 10^{-14}$ \cite{Maggiore:1999vm} ,
and those generated during the PBH collapses 
, $\Omega_{\mr{GW}}h^2 \sim 10^{-13}(f_{\mr{GW}}/10^{-8}\mr{Hz})^{-1}$  \cite{gcGW}.

Note, however, that the actual spectrum of GWs calculated 
from (\ref{spectrum}) has a much larger and sharper peak at 
$f_{\mr{GW}}$ besides the bulk spectrum (\ref{eq:agw}) due to amplification 
caused by resonance between the transfer function of the scalar modes 
and the Green function of the GWs (see Eq.(\ref{eq:endbysp})) \cite{Ananda:2006af}. 
Such amplification, called resonant amplification in \cite{Ananda:2006af}, occurs only if the peak
width, $\Delta$, of the primordial scalar fluctuation is sufficiently
small, $\Delta \ll k_p/2$. Since the resonant growth of the 
amplitude depends on 
the detailed shape of the primordial power spectrum around the peak, 
we do not incorporate it, which yields a conservative bound on the PBH abundance.

We now compare our results with observational constraints.
For definiteness we identify $M_{\mr{PBH}}$ with
the horizon mass when the peak scale $k_p^{-1}$ entered the Hubble
radius.  This is a reasonable approximation even if critical
behavior \cite{critical} is taken into account \cite{massfn}. 
Then $M_{\mr{PBH}}$ is related with the peak frequency of GWs as
\begin{equation}\label{eq:scale}
f_{\mr{GW}} = 0.03~ \mr{Hz} \left(\frac{M_{\mr{PBH}}}
{10^{20} \,\mr{g}}\right)^{-1/2}\!\!
\lmk\frac{g_{\ast p}}{106.75}\rmk^{-1/12}\!\! . 
\end{equation}
The pulsar timing observations are sensitive to GWs 
with $f > 1/T$ where $T$ is the data span. 
Moreover, since pulsars are observed once every few weeks, 
detectable GW frequencies are limited to $f \lesssim 10^{-7}~{\rm Hz}$. Therefore, by using the pulsar timing observations, we can investigate the abundance of PBHs with masses $10^{-2}M_{\odot} \lesssim M_{\mr{PBH}} \lesssim 10^2M_{\odot}(T/35~{\rm yr})^2$.

Space-based laser interferometers are sensitive to 
GWs with $10^{-5}\mr{Hz} \lesssim f \lesssim 10 \mr{Hz}$, which 
covers the entire mass range of the PBHs which are 
allowed to be DM, $10^{20}\mr{g}<M_{\mr{PBH}}<10^{26}\mr{g}$. 
LISA will have its best sensitivity
 $\Omega_{\mr{GW}}h^2 \sim 10^{-11}$ at 
$f \sim 10^{-2} \mr{Hz}~(M_{\mr{PBH}} \sim 10^{21}\mr{g})$, 
BBO and the ultimate-DECIGO are planned to have sensitivities 
$\Omega_{\mr{GW}}h^2 \sim 10^{-13}$ and 
$\Omega_{\mr{GW}}h^2 \sim 10^{-17} $, respectively at 
$f \sim 10^{-1} \mr{Hz}~(M_{\mr{PBH}}\sim10^{19}\mr{g})$
\cite{Sensitivity,Kudoh:2005as}.
 %\begin{widetext}
  \begin{figure}[t]
	\includegraphics[width=1.0\linewidth]{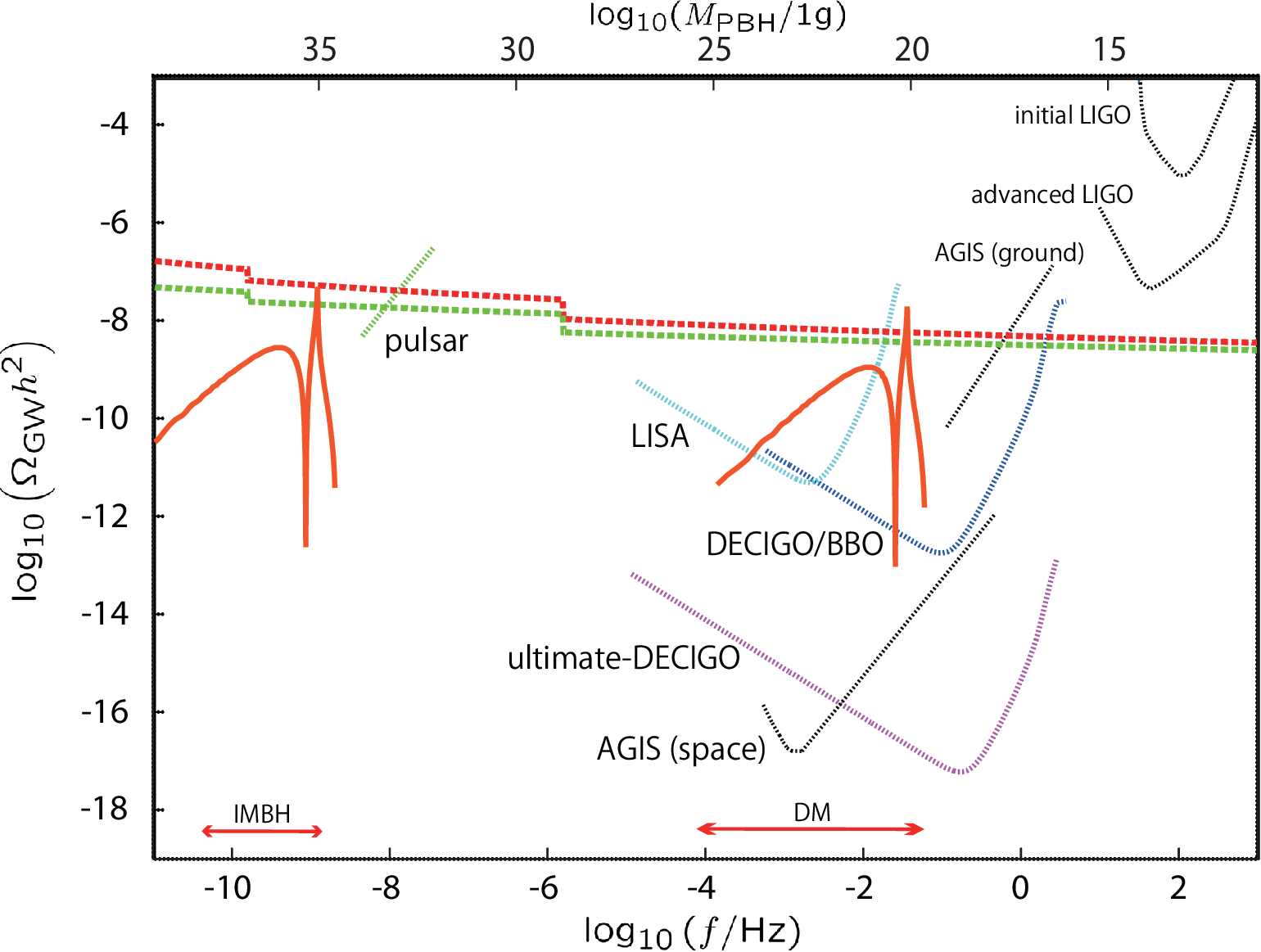}
		\caption{
Energy density of scalar-induced GWs associated with PBH formation 
together with current pulsar constraint (thick solid line segment)
and sensitivity of various GW detectors (convex curves).
Left and right wedge-shaped curves indicate expected power spectra of GWs from 
two different peaked scalar fluctuations corresponding to 
$(\Omega_{\mr{PBH}}h^2,M_{\mr{PBH}},g_{\ast p})
=(10^{-5},30M_{\odot},10.75)$ (left) and
 $(10^{-1},10^{20}\mr{g}, 106.75)$ (right), respectively.
The red dotted (green broken) line shows an envelope curve, $A_{\mr{GW}}$,
 corresponding to $\Omega_{\mr{PBH}}=10^{-1}$ ($10^{-5}$)
obtained by moving $k_p$ and ${\cal A}$, 
which depend on the frequency logarithmically except for
the discontinuities due to the change of the relativistic 
degrees of freedom at the QCD phase transition 
and the electron-positron
pair annihilation.
}
		\label{fig:constraints}
%	\end{figure*}
	\end{figure}
%\end{widetext} 

Figure \ref{fig:constraints} shows the energy density of the
induced GWs obtained by numerically evaluating (\ref{eq:endbysp})
and tracing its subsequent evolution up to the present, whose
peak amplitude is given by  (\ref{eq:agw}).  The left wedge-shaped
curve represents the case $k_p=0.6~\mr{pc}^{-1}$ and $\mc{A}=7 \times 10^{-2}$ corresponding
to $M_{\mr{PBH}}=30\msolar$ and 
$\Omega_{\mr{PBH}}h^2=10^{-6}$, while the right wedge-shaped curve
depicts the case $k_p=2 \times 10^7~\mr{pc}^{-1}$ and $\mc{A}=6 \times 10^{-2}$ corresponding
to $M_{\mr{PBH}}=1 \times 10^{20}$g and 
$\Omega_{\mr{PBH}}h^2=10^{-1}$.  We have also shown the limit imposed
by the pulsar timing observation \cite{Thorsett:1996dr}
and the planned 
sensitivity of space-based laser interferometers
depicted \cite{Sensitivity} with the instrumental parameters used in \cite{Kudoh:2005as}
as well as those of
AGIS \cite{Dimopoulos:2008sv} and LIGO \cite{LIGO}. 

As is seen in the figure the pulsar timing constraint is so stringent
that one cannot achieve $\Omega_{\mr{PBH}}h^2 \geq 10^{-5}$ for
PBHs with $10^{-3}M_{\odot} \lesssim M_{\mr{PBH}} \lesssim M_{\odot}$. 
By observing pulsars for a longer period, we can constrain GWs with lower frequencies, 
which correspond to heavier PBHs. To detect the GWs associated with IMBH-PBHs, we need to observe pulsars for a period, $T \simeq 35~\mr{yr}(M_{\mr{PBH}}/10^{2}M_{\odot})^{\frac{1}{2}}$. Since the GW spectrum extends up to $f=\sqrt{3}f_{\rm GW}$, twenty-years observations 
could detect the GWs corresponding to IMBH-PBHs with masses $M_{\rm PBH} \sim 10^2 M_{\odot}$.

It is clear from Fig.\ \ref{fig:constraints} that 
the future space-based laser interferometers and AGISs can test the feasibility 
of PBHs being the dominant constituent of the DM.
LIGO, on the other hand, has good 
sensitivity at $f \sim 10-10^2 \mr{Hz}$ \cite{LIGO}. 
This frequency band corresponds to mass scale 
$M_{\mr{PBH}} \sim 10^{13}\mr{g}-10^{15}\mr{g}$. 
Though the sensitivity of LIGO is too low now and
in the near future to detect
GWs from the second-order effect associated with 
PBH formation, we could improve the
sensitivity by correlation analysis to reach the
desired level to probe PBHs. Therefore, it may 
be possible to constrain the abundance of the PBHs 
with $M_{\mr{PBH}}<7 \times 10^{14}\mr{g}~(f_{\mr{GW}}>1 
\times 10\mr{Hz})$, which have evaporated by the 
present epoch and could contribute to cosmic rays. 
Further study, however, is necessary in order to 
obtain the conclusion because there are astronomical 
sources of GWs in this frequency band too.  

  \begin{figure}[t]
\includegraphics[width=1.0\linewidth]{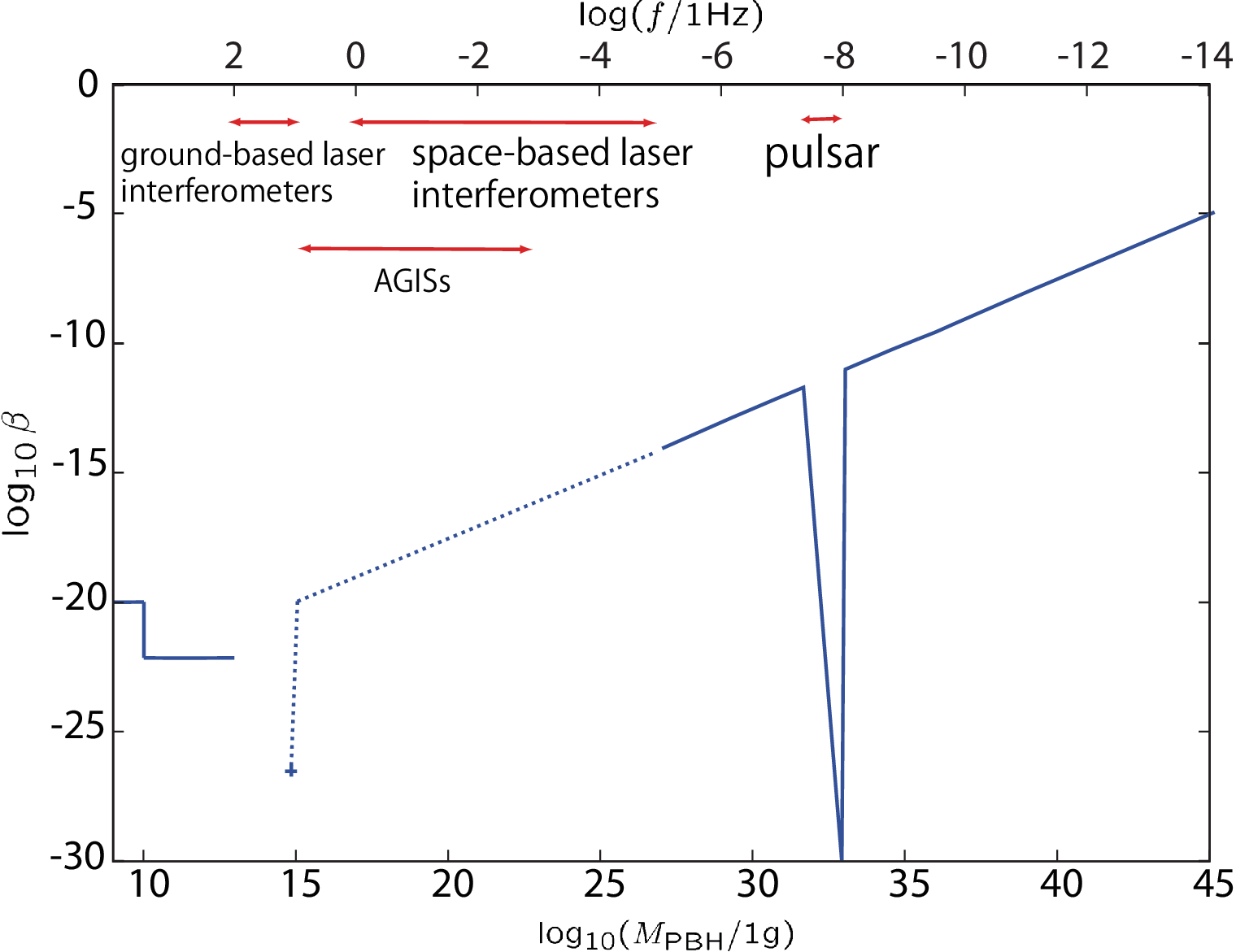}
		\caption{
New constraints on the mass spectrum of PBHs imposed by 
scalar-generated GWs. Dotted line represents the mass range
to be constrained by future GW detectors.}\label{fig2}
\end{figure}

Figure \ref{fig2} depicts the improved constraints on the
PBH fraction $\beta(M_{\mr{PBH}})$ where the dotted region 
denotes the mass range to be constrained by future 
laser interferometers and AGISs.

In summary, we have calculated the spectrum of the 
stochastic gravitational wave background generated
as a second-order effect from scalar perturbations 
which have a spectrum with a high peak to realize
the formation of appreciable numbers of PBHs.
As a result we have found that PBHs with their mass 
corresponding to that of IMBHs could be probed by 
future long-term obsevations of pulsar timing. 
We have also found that if PBHs with mass
$10^{20} - 10^{26}$g are dominant constituents of DM,
we can easily detect the relevant GWs by future
space-based laser interferometers and AGISs.
Thus %scalar-induced
gravitational waves are a new and powerful probe of 
the mass spectrum of PBHs.

%%%%%%%%%%%%%%%%%%%%%%%
%\section*{Acknowledgement}
We thank K.\ Ichiki for useful comments.
This work was supported in part by
JSPS Grant-in-Aid for Scientific 
Research No.\ 19340054(JY) and
by Global COE Program ``the Physical Sciences Frontier", MEXT, Japan.

\end{document}